\documentclass[submission,copyright,creativecommons]{eptcs}
\providecommand{\event}{LFMTP 2018} % Name of the event you are submitting to
\usepackage[dvipsnames]{xcolor}
\usepackage{amsmath,amssymb,amsthm}
\usepackage{breakurl}             % Not needed if you use pdflatex only.
\usepackage{underscore}           % Only needed if you use pdflatex.
\usepackage[textsize=tiny,backgroundcolor=Yellow,linecolor=Brown,bordercolor=Brown]{todonotes}
\usepackage{xparse}
\usepackage{booktabs}
\usepackage{delarray} % for non-messed-up delimiters around arrays
\usepackage{nth}
\usepackage{mathpartir}
\usepackage{cite}

\usepackage{tikz}
\usetikzlibrary{decorations.pathmorphing}
\usetikzlibrary{arrows}

\NewDocumentCommand\RemNode{mm}{
  \colorlet{saved}{.}
  {\tikz[remember picture,baseline=(#1.base)]\node [inner sep=0] (#1) {${\color{saved}#2}$};}
}

\usepackage{mathtools} %for parenthesization, etc.

\DeclarePairedDelimiter\Parens{\lparen}{\rparen}
\DeclarePairedDelimiter\Squares{\lbrack}{\rbrack}
\DeclarePairedDelimiter\Angles{\langle}{\rangle}
\NewDocumentCommand{\deq}{}{\mathrel{\vcentcolon\equiv}}
\mathchardef\rawmathhyphen="2D
\NewDocumentCommand{\mathhyphen}{}{\mathtt{\rawmathhyphen}}

\NewDocumentCommand{\ZZ}{}{\mathbb{Z}}

\newtheorem{theorem}{Theorem}
\theoremstyle{definition}

\theoremstyle{remark}

\NewDocumentCommand{\FormatKwd}{m}{\textbf{\textsf{#1}}}
\NewDocumentCommand{\RedPRL}{}{\texorpdfstring{\FormatKwd{{\color{red}Red}PRL}}{RedPRL}}
\NewDocumentCommand{\Nuprl}{}{\FormatKwd{Nuprl}}
\NewDocumentCommand{\Agda}{}{\FormatKwd{Agda}}
\NewDocumentCommand{\Coq}{}{\FormatKwd{Coq}}
\NewDocumentCommand{\Matita}{}{\FormatKwd{Matita}}
\NewDocumentCommand{\yacctt}{}{\FormatKwd{yacctt}}
\NewDocumentCommand{\cubicaltt}{}{\FormatKwd{cubicaltt}}

\NewDocumentCommand{\dummy}{}{\underline{\hspace{.5em}}}
\NewDocumentCommand{\CTTColor}{m}{{\color{MidnightBlue}#1}}
\NewDocumentCommand{\CTTOp}{m}{\CTTColor{\mathtt{#1}}}
\NewDocumentCommand{\Val}{m}{\CTTColor{#1}\;\mathtt{val}}
\NewDocumentCommand{\Step}{mm}{\CTTColor{#1} \mapsto \CTTColor{#2}}
\NewDocumentCommand{\Steps}{mm}{\CTTColor{#1} \mapsto^* \CTTColor{#2}}
\NewDocumentCommand{\Eval}{mm}{\CTTColor{#1} \Downarrow \CTTColor{#2}}
\NewDocumentCommand{\Bool}{}{\CTTColor{\CTTOp{bool}}}
\NewDocumentCommand{\True}{}{\CTTColor{\CTTOp{true}}}
\NewDocumentCommand{\False}{}{\CTTColor{\CTTOp{false}}}
\NewDocumentCommand{\If}{ggg}{\CTTColor{\CTTOp{if}\IfValueT{#1}\Parens{#1;#2;#3}}}
\NewDocumentCommand{\FunType}{omm}{\CTTColor{\IfValueTF{#1}{\Parens{#1:#2}}{#2}\to#3}}
\NewDocumentCommand{\Lam}{mm}{\CTTColor{\lambda #1. #2}}
\NewDocumentCommand{\App}{mm}{\CTTColor{#1\,#2}}
\NewDocumentCommand{\PairType}{omm}{\CTTColor{\IfValueTF{#1}{\Parens{#1:#2}}{#2}\times#3}}
\NewDocumentCommand{\Pair}{mm}{\CTTColor{\Angles{#1,#2}}}
\NewDocumentCommand{\Fst}{m}{\CTTColor{\CTTOp{fst}\Parens{#1}}}
\NewDocumentCommand{\Snd}{m}{\CTTColor{\CTTOp{snd}\Parens{#1}}}
\NewDocumentCommand{\Circle}{}{\CTTOp{S1}}
\NewDocumentCommand{\Base}{}{\CTTOp{base}}
\NewDocumentCommand{\Loop}{m}{\CTTColor{\CTTOp{loop}_{#1}}}
\NewDocumentCommand{\CircleRec}{gggg}{\CTTColor{\CTTOp{S1\mathhyphen{}rec}\IfValueT{#1}{\Parens{#1;#2;#3;#4}}}}

\NewDocumentCommand{\TypePER}{gg}{\IfValueTF{#1}{\CTTColor{#1}\approx\CTTColor{#2}}{\approx}}
\NewDocumentCommand{\PER}{mgg}{\IfValueTF{#2}{\CTTColor{#2}\mathbin{\approx_{\CTTColor{#1}}}\CTTColor{#3}}{\approx_{\CTTColor{#1}}}}

\NewDocumentCommand{\EqType}{ggo}{\IfValueTF{#1}{\CTTColor{#1}\doteq\CTTColor{#2}\;\IfNoValueTF{#3}{\texttt{type}}{\texttt{type}_{#3}}}{\doteq}}
\NewDocumentCommand{\Type}{mo}{\CTTColor{#1}\;\IfNoValueTF{#2}{\texttt{type}}{\texttt{type}_{#2}}}
\NewDocumentCommand{\Eq}{mmm}{\CTTColor{#2}\doteq\CTTColor{#3}\in\CTTColor{#1}}
\NewDocumentCommand{\Mem}{mm}{\CTTColor{#2}\in\CTTColor{#1}}

\NewDocumentCommand{\Subst}{mmm}{\CTTColor{#1[#2/#3]}}
\NewDocumentCommand{\Hyp}{mm}{\CTTColor{#1}:\CTTColor{#2}}
\NewDocumentCommand{\Seq}{mm}{#1\gg#2}
\NewDocumentCommand{\Cubical}{mm}{#2\;\Squares{\CTTColor{#1}}}

\NewDocumentCommand{\Dim}{m}{\CTTColor{#1}}
\NewDocumentCommand{\DimSubst}{mmm}{\CTTColor{#1\langle#2/#3\rangle}}

\NewDocumentCommand{\InputMode}{m}{{\color{MidnightBlue}{#1}}}
\NewDocumentCommand{\OutputMode}{m}{{\color{Red}{#1}}}
\NewDocumentCommand{\FormatJdg}{m}{{\color{darkgray}{#1}}}

\usepackage{xinttools} % Please replace if you know a better way to do this
\NewDocumentCommand{\FormatList}{mmm}{%
  \xintFor ##1 in {#3} \do{%
    #1{##1}%
    \xintifForLast{}{#2}
  }
}

\NewDocumentCommand{\Rule}{mmm}{
  \begin{array}{l}
    #2\\
    \text{by}\ \text{\FormatKwd{#1}}
    \\
    \;%
    \begin{array}|l.
      \FormatList{}{\\}{#3}
    \end{array}
  \end{array}
}

\NewDocumentCommand{\RedSeqTrue}{mmm}{%
  \FormatJdg{%
    \InputMode{#1}%
    \Longrightarrow%
    \InputMode{#2}\ \mathtt{true}%
    \leadsto%
    \OutputMode{#3}%
  }
}

\NewDocumentCommand{\RedSeqType}{mm}{%
  \FormatJdg{%
    \InputMode{#1}%
    \Longrightarrow%
    \InputMode{#2}\
    \mathtt{type}
  }
}

\NewDocumentCommand{\KindPre}{}{\mathtt{pre}}
\NewDocumentCommand{\KindKan}{}{\mathtt{kan}}

\title{The RedPRL Proof Assistant {\normalsize (Invited Paper)}}
\author{
  Carlo Angiuli
  \institute{Carnegie Mellon University}
  \email{cangiuli@cs.cmu.edu}
  \and
  Evan Cavallo
  \institute{Carnegie Mellon University}
  \email{ecavallo@cs.cmu.edu}
  \and
  Kuen-Bang {Hou (Favonia)}
  \institute{Institute for Advanced Study}
  \email{favonia@math.ias.edu}
  \and
  Robert Harper
  \institute{Carnegie Mellon University}
  \email{rwh@cs.cmu.edu}
  \and
  Jonathan Sterling
  \institute{Carnegie Mellon University}
  \email{jmsterli@cs.cmu.edu}
}
\def\titlerunning{RedPRL}
\def\authorrunning{Angiuli, Cavallo, Hou (Favonia), Harper, Sterling}

\begin{document}
\maketitle

\begin{abstract}
\RedPRL{} is an experimental proof assistant based on Cartesian cubical
computational type theory, a new type theory for higher-dimensional
constructions inspired by homotopy type theory.
In the style of \Nuprl{}, \RedPRL{} users employ tactics to establish behavioral
properties of cubical functional programs embodying the constructive content of
proofs.
Notably, \RedPRL{} implements a two-level type theory, allowing an extensional,
proof-irrelevant notion of exact equality to coexist with a higher-dimensional
proof-relevant notion of paths.
\end{abstract}

\section{Introduction}

Homotopy type theory~\cite{hottbook} and Univalent
Foundations~\cite{voevodsky10hcanon} extend traditional type theory with a
number of axioms inspired by homotopy-theoretic models~\cite{kapulkinlumsdaine},
namely Voevodsky's univalence axiom~\cite{voevodskycmu} and higher inductive
types \cite{lumsdaineshulman}.
In recent years, these systems have been deployed as algebraic frameworks for
formalizing results in synthetic homotopy theory
\cite{coqhott,leanhott,agdahott}, sometimes even leading to the discovery of
novel generalizations of classical theorems \cite{anel2017blakers}.

The constructive character of type theory, embodied in the existence of
canonical forms, is disrupted by axiomatic extensions that are not accounted for
in the traditional computational semantics.
Far from being merely a philosophical concern, the negation of type theory's
native algorithmic content impacts the practice of formalization.
Brunerie proved that the \nth{4} homotopy group of the $3$-sphere is isomorphic
to $\ZZ/n\ZZ$ for some closed $n$, but establishing $n=2$ required years of
additional investigation~\cite{brunerie2016homotopy}.
In ordinary type theory, such a closed $n$ expresses an algorithm that
calculates a numeral; in the absence of computational semantics for homotopy type theory,
there is no reason to expect Brunerie's proof to specify such an algorithm.

Multiple researchers have recently established the constructivity of univalence
and higher inductive types by extending the syntax and semantics of type theory
with cubical machinery.
One solution is embodied in the (De Morgan) cubical type theory of Cohen et
al.~\cite{cohen2018cubical}, implemented in the \cubicaltt{} type
checker~\cite{cubicaltt}.
\RedPRL{} is an interactive proof assistant based on another approach, Cartesian
cubical computational type theory~\cite{ahw2017cubical,afh17uuee,ch18inductive}.

In Section~\ref{sec:computation} we discuss the methodology of computational
type theory, as pioneered in the \Nuprl{} system~\cite{constableetalnuprl}, and
in Section~\ref{sec:proof} we describe its proof-theoretic realization in
\RedPRL{}.
In Section~\ref{sec:cubes} we briefly discuss the cubical machinery present in
Cartesian cubical computational type theory, and its \RedPRL{} implementation.
Finally, in Section~\ref{sec:conclusion} we discuss related and future work.

\section{Computational Type Theory}
\label{sec:computation}

\RedPRL{} and \Nuprl{} are based on a \emph{computation-first} methodology in
which the judgments of type theory range over programs from a programming
language whose syntax and dynamics have already been defined.
This is in contrast with other approaches in which proof terms are inductively
defined and equipped after the fact with a reduction semantics or realizability
model.
One advantage of our methodology is
the ability to incorporate features found in computer science more easily,
such as exceptions and non-termination,
because the theories arise directly from programming languages.
%% Do we have any non-trivial examples of the above? This is the promise and the goal, but it
%% seems like every time we try to import something non-trivial from CS (like state, concurrency, etc)
%% it ends in tears. - JMS
%% Exceptions might be a good example! - JMS
%% And non-termination. Anything we know how to do with logical relations. Do we know how to construct
%% a good logical relation for states? Ask Danny, and I think the answer is still no. -favonia

\begin{figure}[t]
  \begin{align*}
    \text{Expressions}\quad\CTTColor{M,N,O} &\deq
      \CTTColor{x} \mid \FunType[x]{M}{N} \mid \Lam{x}{M} \mid \App{M}{N}
      \\&\ \ \mid
      \PairType[x]{M}{N} \mid \Pair{M}{N} \mid \Fst{M} \mid \Snd{M}
      \\&\ \ \mid
      \Bool \mid \True \mid \False \mid \If{M}{N}{O}
  \end{align*}
  \begin{mathpar}
    \inferrule{ }{\Val{\FunType[x]{M}{N}}}
    \and
    \inferrule{ }{\Val{\Lam{x}{M}}}
    \and
    \inferrule{\Step{M}{M'}}{\Step{\App{M}{N}}{\App{M'}{N}}}
    \and
    \inferrule{ }{\Step{\App{\Parens{\Lam{x}{M}}}{N}}{\Subst{M}{N}{x}}}
    \and
    \inferrule{ }{\Val{\PairType[x]{M}{N}}}
    \and
    \inferrule{ }{\Val{\Pair{M}{N}}}
    \and
    \inferrule{\Step{M}{M'}}{\Step{\Fst{M}}{\Fst{M'}}}
    \and
    \inferrule{ }{\Step{\Fst{\Pair{M}{N}}}{M}}
    \and
    \inferrule{\Step{M}{M'}}{\Step{\Snd{M}}{\Snd{M'}}}
    \and
    \inferrule{ }{\Step{\Snd{\Pair{M}{N}}}{N}}
    \and
    \inferrule{ }{\Val{\Bool}}
    \and
    \inferrule{ }{\Val{\True}}
    \and
    \inferrule{ }{\Val{\False}}
    \and
    \inferrule{\Step{M}{M'}}{\Step{\If{M}{N}{O}}{\If{M'}{N}{O}}}
    \and
    \inferrule{ }{\Step{\If{\True}{N}{O}}{N}}
    \and
    \inferrule{ }{\Step{\If{\False}{N}{O}}{O}}
  \end{mathpar}
  \caption{A programming language with dependent functions, dependent pairs, and booleans.}
  \label{fig:deplang}
\end{figure}

Consider the programming language whose grammar and small-step dynamics are
specified in Figure~\ref{fig:deplang}; the dynamics consists of a stepping
relation $\Step{M}{M'}$ and a value predicate $\Val{\CTTColor{M}}$.
In computational type theory, types are interpreted not as structured sets, but
as \emph{behaviors} specifying classes of programs.
To define such a type theory over this language, we must specify which
programs name types (e.g., $\Bool$), and which programs each type classifies
(e.g., those that evaluate to $\True$ or $\False$).
More precisely, we must consider what types are and when they are equal---in which case they
must classify the same programs---and for each type, what it means for its
elements to evince equal behaviors \cite{cmcp}.

Programs $\CTTColor{A}$ and $\CTTColor{B}$ are equal types, written $\EqType{A}{B}$, when
$\Eval{A}{A_0}$ (i.e., $\Steps{A}{A_0}$ and $\Val{A_0}$), $\Eval{B}{B_0}$, and
$\CTTColor{A_0}$ and $\CTTColor{B_0}$ are equal canonical types.
A canonical type $\CTTColor{A_0}$, in turn, is a value associated to a
(symmetric and transitive) relation on values, specifying the canonical elements
of $\CTTColor{A_0}$ and when two such elements are equal.
Finally, $\CTTColor{M}$ and $\CTTColor{N}$ are equal elements of
$\EqType{A}{A}$, written $\Eq{A}{M}{N}$, when $\Eval{M}{M_0}$, $\Eval{N}{N_0}$,
and $\CTTColor{M_0}$ and $\CTTColor{N_0}$ are classified as equal elements by
$\CTTColor{A_0}$ (where $\Eval{A}{A_0}$).
For notational convenience, we write $\Type{A}$ when $\EqType{A}{A}$, and
$\Mem{A}{M}$ when $\Eq{A}{M}{M}$.

Concretely, the booleans are defined by the following clauses:
\begin{mathpar}
\EqType{\Bool}{\Bool} \and
\Eq{\Bool}{\True}{\True} \and
\Eq{\Bool}{\False}{\False}
\end{mathpar}
That is, $\Bool$ is a canonical type with canonical (unequal) elements $\True$
and $\False$.
The canonicity property follows directly from the above definitions.

\begin{theorem}[Canonicity]
  For any $\Mem{\Bool}{M}$, $\Eval{M}{\True}$ or $\Eval{M}{\False}$.
\end{theorem}

The judgments for programs with free variables are defined in terms of all
closing substitutions.
In the case with one variable, we say $\Seq{\Hyp{x}{A}}{\EqType{B}{C}}$ when for
all $\CTTColor{M}$ and $\CTTColor{N}$, if $\Eq{A}{M}{N}$ then
$\EqType{\Subst{B}{M}{x}}{\Subst{C}{N}{x}}$.
That is, the type families $\CTTColor{B}$ and $\CTTColor{C}$ are equal when they
send equal elements of $\CTTColor{A}$ to equal types.
Similarly, we say $\Seq{\Hyp{x}{A}}{\Eq{B}{M}{N}}$ when for all $\CTTColor{O}$
and $\CTTColor{P}$, if $\Eq{A}{O}{P}$ then
$\Eq{\Subst{B}{O}{x}}{\Subst{M}{O}{x}}{\Subst{N}{P}{x}}$.
Dependent function and dependent pair types are then defined by the following
clauses:
\begin{align*}
  \EqType{\FunType[x]{A}{B}}{\FunType[x]{C}{D}}
    &\iff \EqType{A}{C}\ \text{and}\ \Seq{\Hyp{x}{A}}{\EqType{B}{D}} \\
  \Eq{\FunType[x]{A}{B}}{\Lam{x}{M}}{\Lam{x}{N}}
    &\iff \Seq{\Hyp{x}{A}}{\Eq{B}{M}{N}} \\
  &\\[-.8em]
  \EqType{\PairType[x]{A}{B}}{\PairType[x]{C}{D}}
    &\iff \EqType{A}{C}\ \text{and}\ \Seq{\Hyp{x}{A}}{\EqType{B}{D}} \\
  \Eq{\PairType[x]{A}{B}}{\Pair{M}{N}}{\Pair{O}{P}}
    &\iff \Eq{A}{M}{O}\ \text{and}\ \Eq{\Subst{B}{M}{x}}{\Subst{N}{M}{x}}{\Subst{P}{O}{x}}
\end{align*}

In general, new type constructors are implemented by extending the syntax and
dynamics of the programming language, and adding clauses to the definitions of
canonical types and elements.
For the full programming language currently in \RedPRL{}
and the technical details of the defining clauses, see
Angiuli et al.~\cite{afh17uuee} and Cavallo and Harper~\cite{ch18inductive}.

Open judgments in computational type theory are distinct
from those in many type theories:
variables range over closed programs,
instead of being indeterminate objects (sometimes called ``generic values'').
One consequence is that computational type theory often validates
relatively strong extensionality principles, including the full universal property of the $\Bool$ type
(for all terms $\Eq{\Bool}{M}{M}$ and $\Seq{\Hyp{x}{\Bool}}{\Eq{C}{N}{N}}$):
\[
  \Eq{
    \Subst{C}{M}{x}
  }{
    \Subst{N}{M}{x}
  }{
    \If{M}{
      \Subst{N}{\True}{x}
    }{
      \Subst{N}{\False}{x}
    }
  }
  % \Seq{\Hyp{x}{\Bool}}{\Eq{\mathcal{F}[x]}{\mathcal{E}[x]}{\mathcal{E}[\If{x}{\True}{\False}]}}
\]
(This equation is called the \emph{Shannon expansion}, a widely used tool for deciding boolean
formulae.)

The above cannot be established as a judgmental equality in most type theories,
and while it is known how to implement extensional booleans in a
proof-theoretic setting (at least in the simply-typed case), the same
techniques cannot be applied to inductive types with recursive generators (such
as natural numbers), nor even to the empty coproduct if regarded as a positive
type.\footnote{The universal property of the positive version of the empty type
entails a collapse of definitional equivalence in an inconsistent context; in a
traditional proof-theoretic account of type theory, this disrupts strong
normalization.} In computational type theory, however, all of these unicity
principles, as well as function extensionality, follow immediately from the
definitions of the open judgments and equality for each type.

%This includes $\If{\True}{\False}{O}$
%for any closed $\CTTColor{O}$ because it will evaluate to $\False$.

%A function type $\FunType{A}{B}$ should recognize any computable function
%that evaluates to a program of form $\Lam{x}{M}$ and
%sends equal elements in $\CTTColor{A}$ to equal elements $\CTTColor{B}$.
%Two functions are the same if they give equal results on the same inputs.

\section{Proof Refinement Logic}
\label{sec:proof}

It is crucial, after defining a computational type theory from a programming language, to
devise a proof theory that is sound and facilitates formalization and computer
checking.  Proof theories prioritize usability and are often incomplete; as
such, we have been experimenting with multiple different designs in parallel
with \RedPRL{} to find out what works the best for our intended applications;
\RedPRL{} is our first such experiment, but there are many other possibilities
within the design space~\cite{cubicaltt}.

For example, some proof theories achieve a decision procedure for
type checking by including more data in the terms, at the expense of making
certain kinds of equational reasoning more baroque; this approach has some
undeniable advantages in comparison to \RedPRL's more extrinsic style, and we
are currently studying it as a means to alleviate a number of practical
difficulties in the \RedPRL{} system.  We do not at this time recognize any
particular proof theory as the ``canonical'' one.

The proof theory in \RedPRL{} is a \emph{proof (program) refinement logic} in the tradition of
\Nuprl{}~\cite{constableetalnuprl}; as such, it is oriented around the
decomposition of proof goals into subgoals, and the extraction of computational
evidence for main goals from the evidence for the subgoals, a refinement of the
LCF methodology first synthesized by~\cite{bates:1981}.
% I am not sure if the above is clear: I want to say that Bates synthesized
% this refinement of LCF, not that Bates invented LCF.
% favonia: how about just breaking the sentence into two parts, having Bates
% as the subject of the second one?
At a technical level, the \RedPRL{} proof logic is an extraction-oriented
sequent calculus with judgments like the following, which asserts that the type
$\InputMode{A}$ is inhabited under the assumptions $\InputMode{H}$, with
realizer or witness $\OutputMode{M}$:
\[
  \RedSeqTrue{H}{A}{M}
\]

In the above, $\InputMode{H}$ and $\InputMode{A}$ are inputs to the judgment,
whereas $\OutputMode{M}$ is an \emph{output} to the judgment: this means that
it does not appear in the statement of the goal, but is instead synthesized by
the proof refinement system in the course of the proof. A proof refinement
logic for this judgment is a signature of \emph{refinement rules}, which
explain how to decompose one such sequent into a collection of other sequents
whose witnesses can be combined into a witness for the main sequent.

In \RedPRL, these refinement rules are written downward as in the following
example:
\[
  \Rule{sigma/intro}{
    \RedSeqTrue{H}{
      \Parens{x:A}\times B(x)
    }{
      \Angles{
        \RemNode{Tuple/M}{M},
        \RemNode{Tuple/N}{N}
      }
    }
  }{
    \RedSeqTrue{H}{A}{\RemNode{M}{M}},
    \RedSeqTrue{H}{B(\RemNode{N/aux}{M})}{\RemNode{N}{N}},
    \RedSeqType{H,x:A}{B(x)}
  }
\]
\begin{tikzpicture}[overlay,remember picture, very thick]
  \draw [->, MidnightBlue] (M.west) to [out=90,in=90] (N/aux.north);
  \draw [->, Red] (M.east) to [out=0, in=-30] (Tuple/M.south);
  \draw [->, Red] (N.east) to [out=0, in=-30] (Tuple/N.south);
\end{tikzpicture}

Above, the binding structure of the rule is indicated with arrows; in general,
\OutputMode{outputs} flow downward through the subgoals and upward to the
\OutputMode{output} of the main goal.
As can be seen, our schema for refinement rules generalizes the one used in
\Nuprl{}, in that it permits the statements of subgoals to depend on the
realizers of earlier subgoals~\cite{sterling2017}; this facility also appears
in the latest version of \Coq's refinement engine~\cite{spiwack2011}, and is
also present in the \Matita{} proof assistant~\cite{matita2011}.

The complete collection of \RedPRL{}'s refinement rules can be viewed
online~\cite{redprl}.

\paragraph{Auxiliary subgoals}

In addition to the two familiar subgoals to the introduction rule for dependent
pairs, there is a third subgoal required to establish that $B(x)$ is
actually a genuine family of types indexed in $A$. In proof assistants like
\Agda{} and \Coq{}, these kinds of obligations are checked automatically;
however, \RedPRL{} is based on realizers rather than checkable proof terms, so
we must explicitly emit such a subgoal in order to preserve soundness. In
nearly all cases, such auxiliary subgoals can be discharged automatically using
the \FormatKwd{auto} tactic.

\subsection{Rule Composition and Proof Tactics}

A proof in \RedPRL{} is built from
\emph{tactics}~\cite{gordon-milner-wadsworth:1979}; every refinement rule is a
tactic, but the tactics include composite forms like the following:

\begin{center}
  \begin{tabular}{ll}
    \toprule
    \textit{Expression} & \textit{Meaning}
    \\
    \midrule
    $\mathtt{t_1; t_2}$
    & ``run $\mathtt{t_1}$ on the current goal, and then run $\mathtt{t_2}$ on the resulting subgoals''
    \\
    $\mathtt{t; \Squares{t_0,\ldots,t_n}}$
    & ``run $\mathtt{t}$ on the current goal, and then run $\mathtt{t_i}$ on the $i$th resulting subgoal''
    \\
    $\mathtt{t_1\mid{}t_2}$
    & ``try running $\mathtt{t_1}$ on the current goal, but if that fails, run $\mathtt{t_2}$ instead''
    \\
    $\FormatKwd{auto}$
    & ``use automation to decompose the current goal as much as possible''
    \\
    \ldots
    \\
    \bottomrule
  \end{tabular}
\end{center}

In order to ease the process of constructing proofs and programs in \RedPRL{},
the tactic language is also equipped with special notation for certain frequent
combinations of rules and tactics. For instance, the following tactic
introduces two variables and pairs them, solving the goal
$\Parens{x:A}\to\Parens{y:B(x)}\to\Parens{x:A}\times B(x)$:
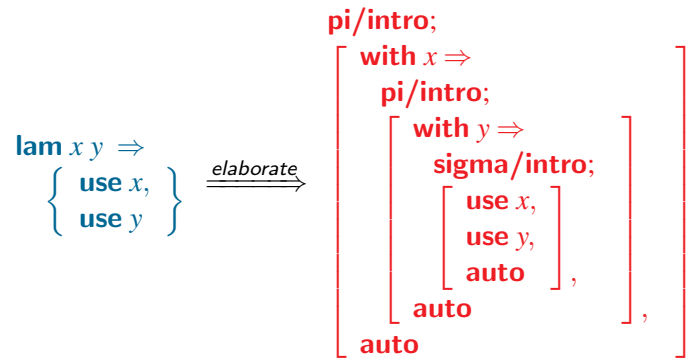
\[
  \InputMode{
    \begin{array}[c]{l}
      \FormatKwd{lam}\ x\ y\ \Rightarrow
      \\
      \quad
      \begin{array}[t]\{{l}\}
        \FormatKwd{use}\ x,\\
        \FormatKwd{use}\ y
      \end{array}
    \end{array}
  }
  \xRightarrow{
    \textit{\textsf{elaborate}\ }
  }
  \OutputMode{
    \begin{array}[c]{l}
      \FormatKwd{pi/intro};\\
        \begin{array}\lbrack{l}\rbrack
          \FormatKwd{with}\ x \Rightarrow\\
          \
          \begin{array}{l}
            \FormatKwd{pi/intro};\\
            \begin{array}[b]\lbrack{l}\rbrack
              \FormatKwd{with}\ y \Rightarrow\\
              \
              \begin{array}{l}
                \FormatKwd{sigma/intro};\\
                \begin{array}[b]\lbrack{l}\rbrack
                  \FormatKwd{use}\ x,\\
                  \FormatKwd{use}\ y,\\
                  \FormatKwd{auto}
                \end{array},
              \end{array}\\
              \FormatKwd{auto}
            \end{array},
          \end{array}
          \\
          \FormatKwd{auto}
        \end{array}
    \end{array}
  }
\]

While the tactic notation above is built-in for convenience, users can also
define their own tactics and tacticals, though \RedPRL{} does not yet provide
any facilities for extending the concrete notation.

\section{Cubical Type Theory}
\label{sec:cubes}

The theory underlying \RedPRL{} differs from earlier computational type theories
(notably \Nuprl{}) by extending the syntax of the programming language with \emph{dimension
expressions} encoding higher-dimensional path structure.
A dimension expression $\Dim{r}$ is either a dimension name $\Dim{i}$
(representing a generic element of an abstract interval), or a constant $\Dim0$
or $\Dim1$ representing one of the two endpoints.

In Cartesian cubical computational type theory,
there is always an ambient context $\CTTColor{\Psi}$ of dimension names,
signifying the dimensions accessible to the program.
Dimension names behave like variables because
one can substitute a dimension expression for a name,
where a $\Dim0$- or $\Dim1$-substitution takes the $\Dim0$- or $\Dim1$-face, respectively,
and substituting one name for another takes the corresponding diagonal.

For example, let $\CTTColor{M}$ be a program indexed by dimension $\Dim{i}$,
$\DimSubst{M}{0}{i}$ is its $\Dim0$-face, $\DimSubst{M}{1}{i}$ is its $\Dim1$-face,
and $\DimSubst{M}{j}{i}$ is the diagonal identifying dimensions $\Dim{i}$ and $\Dim{j}$.
A program indexed by $n$ dimension names (i.e., by an $n$-fold \emph{Cartesian
product} of abstract intervals) forms an abstract $n$-dimensional \emph{cube}.
Using dimension names, it is possible to express higher-dimensional structure at
the judgmental level.
For example, the judgment $\Cubical{\Dim{i}}{\Eq{A}{M}{N}}$
means that $\CTTColor{M}$ and $\CTTColor{N}$ are equal \emph{lines}, or paths,
varying in the dimension $\Dim{i}$.
In general, the judgements $\Cubical{\Psi}{\EqType{A}{B}}$
and $\Cubical{\Psi}{\Eq{A}{M}{N}}$
refer to $|\CTTColor{\Psi}|$-dimensional structure varying in the dimensions $\CTTColor{\Psi}$.
(See Angiuli et al.~\cite{ahw2017cubical} for a more detailed introduction to
cubical type theory.)

\subsection{Stable Computation}
\label{sec:stable-comp}

\begin{figure}[t]
  \begin{align*}
    \text{Dimension expressions}&& \CTTColor{r} &\deq \Dim0 \mid \Dim1 \mid \Dim{i}
    \\
    \text{Expressions}&& \CTTColor{M,N,O,P} &\deq
      \cdots
      \mid \Circle \mid \Base \mid \Loop{r} \mid \CircleRec{x.M}{N}{O}{i.P}
  \end{align*}
  \begin{mathpar}
    \inferrule{ }{\Val{\Circle}}
    \and
    \inferrule{ }{\Val{\Base}}
    \and
    \inferrule{ }{\Val{\Loop{i}}}
    \and
    \inferrule{ }{\Step{\Loop{0}}{\Base}}
    \and
    \inferrule{ }{\Step{\Loop{1}}{\Base}}
    \and
    \inferrule{\Step{M}{M'}}{\Step{\CircleRec{x.C}{M}{B}{i.L}}{\CircleRec{x.C}{M'}{B}{i.L}}}
    \and
    \inferrule{ }{\Step{\CircleRec{x.C}{\Base}{B}{i.L}}{B}}
    \and
    \inferrule{ }{\Step{\CircleRec{x.C}{\Loop{j}}{B}{i.L}}{\DimSubst{L}{j}{i}}}
  \end{mathpar}
  \caption{A fragment of the circle type, omitting the Kan structure.}
  \label{fig:circlelang}
\end{figure}

Dimension substitutions significantly complicate the story of computational type
theory, because they do not in general commute with computation.

To see the problem,
let's introduce a higher inductive type representing a circle, described in part in
Figure~\ref{fig:circlelang}; the new constructs include:
\begin{itemize}
  \item $\Circle$: the circle type.
  \item $\Base$: a point constructor in $\Circle$, representing a distinguished point in the circle.
  \item $\Loop{r}$: a path constructor parametrized by the dimension expressions $\Dim{r}$,
    representing the loop in the circle.
    The program that $\Loop{i}$ represents is a line along dimension $\Dim{i}$ from $\Base$ to $\Base$,
    which is witnessed by the computation rules $\Step{\Loop{0}}{\Base}$ and $\Step{\Loop{1}}{\Base}$.
  \item $\CircleRec{x.C}{M}{B}{i.L}$: the eliminator of $\Circle$ by case analysis,
    where $\CTTColor{x.C}$ is the motive, $\CTTColor{M}$ is the target,
    and $\CTTColor{B}$ and $\CTTColor{i.L}$ are the two methods matching the constructors $\Base$ and $\Loop{i}$.
\end{itemize}

With the circle type $\Circle$, we can see how dimensions might complicate the story of computation.
Consider the program $\CircleRec{\dummy.\Bool}{\Loop{i}}{\True}{\dummy.\False}$.
It should evaluate to $\False$ according to the rules in Figure~\ref{fig:circlelang}.
However, if we substitute $\Dim0$ for $\Dim{i}$ before evaluating, the resulting
program instead evaluates to $\True$:
\begin{align*}
  \CircleRec{\dummy.\Bool}{\Loop{0}}{\True}{\dummy.\False}
  &\mapsto
  \CircleRec{\dummy.\Bool}{\Base}{\True}{\dummy.\False}
  \\
  &\mapsto
  \True.
\end{align*}

This is a serious issue if we hope to close our type theory under a
computational notion of equivalence.  We therefore must restrict our
type theory to only recognize programs for which computation and dimension
substitution commute (up to judgmental equality); this essential restriction
ensures that the program above shall not inhabit the $\Bool$ type.
We also identify a class of \emph{stable} computation rules that always commute with
dimension substitution, and therefore can be used to simplify programs without
first establishing that they are well-typed.

Angiuli et al.~\cite{afh17uuee} identify many computation rules as stable: all rules present in
traditional type theory, as they are defined uniformly across all dimension contexts; moreover,
rules such as $\Step{\Loop{0}}{\Base}$ will never be affected by substitutions.  \RedPRL{} is
equipped with an algorithm to identify more subtle instances of stable computation, by taking into
account which dimensions are bound and thus unaffected by substitutions.  However, our experience
shows that this algorithm does not help as much in practice as one might hope: many difficult cases
that arise in concrete formalization efforts are not precisely stable, rather only up to typed
judgmental equality.

On the other hand, programs that are already known to be well-typed admit a much broader collection
of computation rules. We believe that by focusing our efforts in the future on developing an
ergonomic theory of typed programs, we can employ a richer version of definitional equivalence
including equations not justified by stable computation alone.

\subsection{Kan Operations and Kinds}

In order to perform the homotopy-theoretic constructions available in homotopy
type theory, we must equip types with \emph{Kan operations} that describe how to
compose and invert lines, and how to transport elements along lines of
types~\cite{ahw2017cubical,afh17uuee}.
Types $\CTTColor{A}$ equipped with such operations are called Kan, and written
$\Cubical{\Psi}{\Type{A}[\KindKan]}$.
They are also called \emph{fibrant types} because the fibrancy
in many models of interest corresponds to implementing Kan operations.
All \RedPRL{} type formers present in homotopy type theory---including dependent
functions, dependent pairs, paths, higher inductive types, and type
universes---preserve being Kan.

However, \RedPRL{} also contains types \emph{not} present in homotopy type
theory, notably, \emph{exact equality types} internalizing judgmental equality.
These cannot in general be equipped with Kan operations, because equality is finer than
homotopy and therefore not invariant under paths.
Such \emph{pretypes} $\CTTColor{A}$ are written $\Cubical{\Psi}{\Type{A}[\KindPre]}$.

Pretypes and Kan types are part of a continuum of Kan structures recognized by
\RedPRL{}, such as the types with only partial Kan structure, or those with
stronger properties (such as having no non-trivial paths).
We are able to further refine our type theory with a language of (currently,
five) \emph{kinds} $\kappa$ governing the type equality judgment
$\Cubical{\Psi}{\EqType{A}{B}[\kappa]}$, resulting in a theory more expressive than other
two-level type theories such as~\cite{ack17twolevel}.
Equality types of a type with no non-trivial paths can be made Kan, for example.

\section{Related and Future Work}
\label{sec:conclusion}

\RedPRL{} is under active development.
We are currently implementing support for general higher inductive
types~\cite{ch18inductive}, which will greatly expand the range of
homotopy-theoretic proofs expressible in \RedPRL{}.
One consequence would be the ability to fully model homotopy type theory
and various two-level type theories\footnote{Except the resizing rules in the homotopy type system proposed by Voevodsky~\cite{voevodsky13hts}.};
see \cite{ch18inductive}.
We are also actively experimenting with alternate implementations of complex Kan
operations (in particular, for the $\CTTColor{\CTTOp{V}}$ and
$\CTTColor{\CTTOp{Fcom}}$ types in Angiuli et al.~\cite{afh17uuee}).

The use of the cubical structure in constructive models of homotopy type theory was
pioneered by Bezem et al.~\cite{bch}, and has since been studied extensively by
ourselves and
others~\cite{licata2014cubical,cohen2018cubical,ahw2017cubical,afh17uuee,ch18inductive,abcfhl17cart}.
There are many subtle technical choices possible in this arena---such as the
tradeoffs between expressivity of the dimension language and Kan
operations---but these are outside the scope of this survey article.

With Anders M\"ortberg, we are also implementing the \yacctt{} type checker for
a variant of Cartesian cubical type theory whose formal properties mirror those
of \cubicaltt{}, rather than the proof refinement logic of \RedPRL{}.
In connection with this project, we are studying the question of normal forms
and algorithmic definitional equivalence for Cartesian cubical type theory.
Despite these efforts, the natural number $n$ in $\ZZ/n\ZZ$ mentioned in the
introduction remains to be computed successfully: attempts in \cubicaltt{} have
consumed excessive memory and failed to terminate, while \RedPRL{} and \yacctt{}
do not yet contain sufficient homotopy-theoretic results.

%and tactics for applying unstable computation rules (see
%Section~\ref{sec:stable-comp}).

\section*{Acknowledgements}

We would like to thank everyone who has contributed to the \RedPRL{} project
and its predecessors, including Eugene Akentyev, Tim Baumann, David Thrane
Christiansen, Daniel Gratzer, Darin Morrison, Anders M\"ortberg, Vincent Rahli and James Wilcox.

This research was sponsored by
the National Science Foundation
under grant number DMS-1638352
and
the Air Force Office of Scientific Research
under grant number FA9550-15-1-0053.
The second author would also like to thank the Isaac Newton Institute for Mathematical Sciences
for its support and hospitality during the program ``Big Proof''
when part of work on this paper was undertaken;
the program was supported by Engineering and Physical Sciences Research Council
under grant number EP/K032208/1.
The views and conclusions contained in this document are those of the authors
and should not be interpreted as representing the official policies, either expressed or implied, of any
sponsoring institution, government or any other entity.

\nocite{mcbride:2004, brady:2013}
\bibliographystyle{eptcs}
\bibliography{redprl}
\end{document}